# Van der Waals heteroepitaxial growth of monolayer Sb in puckered honeycomb structure

Zhi-Qiang Shi[1]†, Huiping Li[2,3]†, Qian-Qian Yuan[1], Ye-Heng Song[1], Yang-Yang Lv[4], Wei Shi[1], Zhen-Yu Jia[1], Libo Gao[1], Y. B. Chen[1], Wenguang Zhu[2,3]*, Shao-Chun Li[1,5]*

[1]*National Laboratory of Solid State Microstructures, School of Physics, Nanjing University, Nanjing 210093, China*

[2]*International Center for Quantum Design of Functional Materials (ICQD), Hefei National Laboratory for Physical Sciences at the Microscale, and Synergetic Innovation Center of Quantum Information and Quantum Physics, University of Science and Technology of China, Hefei, Anhui 230026, China*

[3]*Key Laboratory of Strongly-Coupled Quantum Matter Physics, Chinese Academy of Sciences, School of Physical Sciences, University of Science and Technology of China, Hefei, Anhui 230026, China*

[4]*National Laboratory of Solid State Microstructures, Department of Materials Science and Engineering, Nanjing University, Nanjing 210093, China*

[5]*Collaborative Innovation Center of Advanced Microstructures, Nanjing University, Nanjing 210093, China*

† Z.Q.S. and H.L. contributed equally to this work.
* Email: wgzhu@ustc.edu.cn (W.G.Z); scli@nju.edu.cn (S.C.L.)

Abstract: Atomically thin two-dimensional (2D) crystals have gained tremendous attentions owing to their potential impacts to the future electronics technologies, as well as the exotic phenomena emerging in these materials. Monolayer of α phase Sb (α-antimonene) that shares the same puckered structure as black phosphorous, has been predicted to be stable with precious properties. However, the experimental realization still remains challenging. Here, we successfully grow high-quality monolayer α-antimonene, with the thickness finely controlled. The α-antimonene exhibits great stability upon exposure to air. Combining scanning tunneling microscope, density functional theory calculations and transport measurement, it is found that the electron band crossing the Fermi level exhibits a linear dispersion with a fairly small effective mass, and thus a good electrical conductivity. All of these properties make the α-antimonene promising in the future electronic applications.

Spurred by their prospect in electronic technologies, two-dimensional (2D) crystals have been attracted increasing attentions. As the thickness is decreased down to the single-layer limit, 2D crystals usually exhibit different electronic properties from their bulk counterparts.[1] Exotic phenomena are also expected in single-layered materials, such as the quantum spin Hall effect,[2-6] 2D superconductivity,[7,8] charge density wave,[9-12] or magnetism.[13,14] Following the discovery of graphene,[15,16] black phosphorus (BP) has been revived as a potential candidate for optoelectronics and FET applications,[17-20] which however is suffered from the chemical instability. Many mono-element single-layered materials have been predicted and realized, such as Si,[21-23] Ge,[24,25] Sn,[26] B,[27,28] Hf,[29] and Te,[30] but few of them share the same crystal structure as BP.

It is believed that BP structured monolayer (α-allotrope) can be formed in other group V elements, such as Bi (bismuthene), Sb (antimonene) or As (arsenene), and many theoretical efforts have been made to predict their structures and properties.[31-37]. Comparing to their β-allotrope of hexagonal honeycomb structure that have been widely studied experimentally,[38-44] it still remains challenging to fabricate the large-scale and high-quality monolayer α-allotrope of these group V mono elements, [36] even though small patches of the α-allotrope has been observed in some mixed structures. [45]

In this study, we successfully synthesize the large-scale and high-quality α-antimonene with puckered BP structure on the $T_d$-WTe$_2$ substrate, by using molecular beam epitaxy (MBE). In our experiment, the thickness of BP-structured antimonene can be well controlled in a layer-by-layer fashion. Owing to the high quality and large scale of the Sb monolayer, it becomes possible to map the electronic band structure via quasiparticle interference (QPI) with scanning tunneling microscopy (STM). The α-antimonene exhibits a hole-doped nature with a linearly dispersed band crossing the Fermi level and a high electrical conductivity. Moreover, the α-antimonene is ultra-stable upon exposure to air, thus making it a good complementary to the family of 2D materials, and should be of potential to many applications.

Free standing α-antimonene takes a distorted BP structure, consisting of two sub-atomic layers, as illustrated in **Figure 1**a. The antimonene layers are stacked along the vertical direction by van der Waals interaction to form the bulk antimony.[34] To achieve the nearly free-standing epitaxial α-antimonene monolayer, the substrate is required to be chemically inert and weakly bonded to suppress the hybridization to the epitaxial layer. $T_d$-WTe$_2$ is a suitable substrate to grow α-antimonene because 1) the surface is relatively inert and 2) the rectangular unit cell of $T_d$-WTe$_2$ surface is compatible to that of α-antimonene, as shown in Figure 1a.

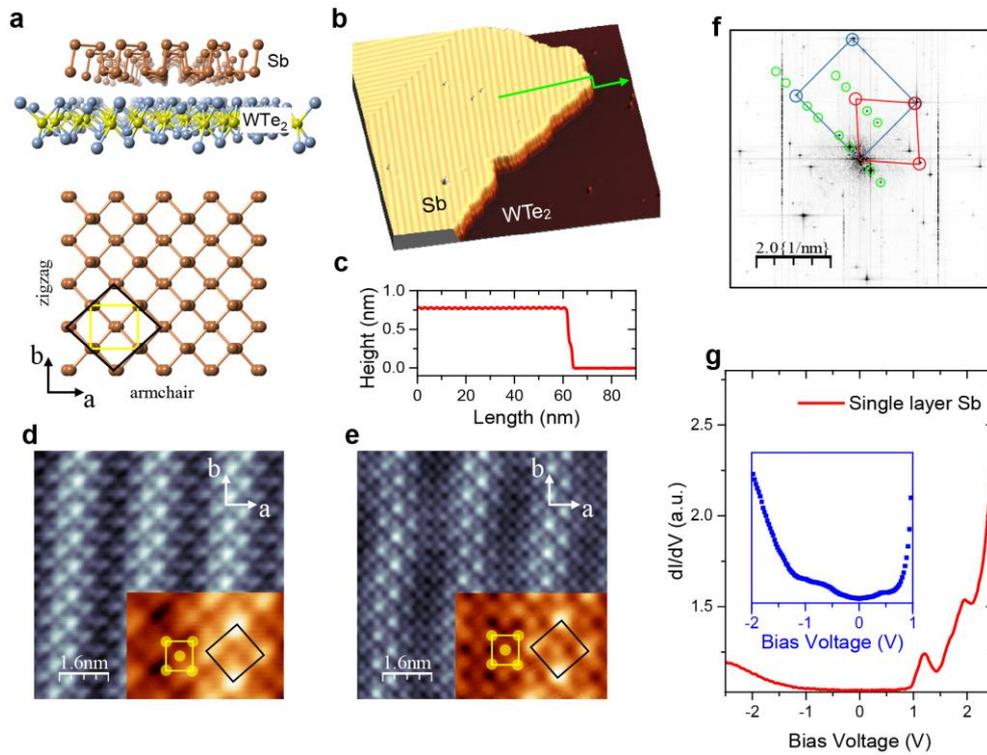

**Figure 1**. STM and STS results of the epitaxial single layered α-antimonene grown on $T_d$-WTe$_2$. a) The upper is the schematic of epitaxial α-Sb on $T_d$-WTe$_2$. The lower is the top view of freestanding α-antimonene, the 1×1 and R2 unit cells are marked by yellow and black rectangles, respectively. The armchair and zigzag directions are labeled. b) The STM topology image (120 × 120 nm$^2$) of single layer α-Sb fabricated on WTe$_2$. $U$ = +2 V, $I_t$ = 100 pA. c) The line-scan profile taken along the green arrowed line in (b). d,e) Atomically resolved STM images (8 × 8 nm$^2$) taken at $U$ = +300 mV and $U$ = -50 mV, respectively. $I_t$ = 100 pA. In the inset, the black and yellow rectangles mark the unit cells of R2 reconstruction and 1×1 lattice. f) FFT of the large-scale STM topology image, the blue, red and green circles mark the Bravais vectors of 1×1, R2 reconstruction and moiré pattern respectively. g) The red curve shows the experimental STS of single layer α-Sb from -2.5 V to +2.5 V. The blue curve is the enlarged part from -2.0 V to +1.0 V.

Initially, Sb on the $T_d$-$WTe_2$ surface prefers a monolayer growth mode. Figure 1b shows the topographic image taken on a surface with ~0.3 ML Sb on the $T_d$-$WTe_2$, where Sb forms the atomically flat islands with no detectable defects. The step height of the Sb island, as plotted in Figure 1c, is ~0.78 nm, slightly larger than the calculated thickness of 0.61 nm for the single-layer antimonene film.[34] A moiré pattern, can be identified in Figure 1b as the ridge-like feature with a period of ~2.5 nm, which is ascribed to the interference between the subtly different lattice periodicity of antimonene and $T_d$-$WTe_2$. Through optimizing the growth parameters, the single-layered film of antimonene can be grown as large as a micrometer, (see Figure S1, Supporting Information). The single-layered island is usually composed of two domains, with the orientation of the moiré pattern rotated by ~80° with respect to each other, (as shown in Figure S2, Supporting Information). The atomic-resolution STM images (Figure 1d,e), as well as the Fast Fourier Transform (Figure 1f), clearly reveal a reconstructed $\sqrt{2}\times\sqrt{2}$ (R2) lattice as marked by the black squares, and the original 1×1 lattice as marked by the yellow squares. The a and b axis of the 1×1 unit cell is experimentally determined to be ~0.44 nm and ~0.48 nm, very close to the calculated values for the free standing monolayer Sb.[34] However, the original rectangular lattices are slightly distorted with an angle of ~91°. In both domains, the moiré pattern is nearly parallel to the armchair direction of the antimonene lattice. It is worthwhile noting that the lattice constants for the $\sqrt{2}\times\sqrt{2}$ unit cell of α-antimonene is 0.62 nm × 0.63 nm, closely compatible with the size of two unit cells of the substrate $WTe_2$, i.e., 0.63 nm × 0.68 nm, (see Figure S3, Supporting Information). Based on these measurements, a structural model of a $\sqrt{2}\times\sqrt{2}$ reconstructed antimonene on $WTe_2$ is proposed (as shown in Figure S2, Supporting Information).

The dI/dV spectra, which represent the local density of state (LDOS), are taken on the terrace of the single-layer antimonene, and shown in Figure 1g. The finite density at the Fermi energy indicates a metallic nature. In fact, the dI/dV spectra are rather uniform along the surface (see Figure S4, Supporting Information), suggesting the high film quality and homogeneous electronic band structure. Furthermore, the dI/dV spectra taken on different domains shows subtle difference as well, (as shown in Figure S4, Supporting Information). By increasing the amount of Sb, multilayered antimonene of large scale can be grown, as shown in **Figure 2**a. Figure 2b-d shows the atomic resolution images taken on the different layers of antimonene, confirming the BP-like puckered structure for all the thickness. The step heights measured on the antimonene films from $1^{st}$ to $6^{th}$ layer are plotted in Figure 2e (see also Figure S5, Supporting Information). The step heights from the $2^{nd}$ to the $6^{th}$ layer are kept nearly constant varying from ~0.65 nm to ~0.63 nm, but smaller than the height of the $1^{st}$ layer, indicating again the weakly coupling between antimonene and the $T_d$-WTe$_2$ substrate. dI/dV spectra are also measured on these multiple layers, as plotted together in Figure 2f. All the dI/dV spectra exhibit the same metallic nature. The characteristic peaks in the unoccupied states, as marked by the black triangles in Figure 2f, gradually shift towards the Fermi energy as the thickness increases.

Raman characterization is performed on the α-antimonene films, as shown in Figure 2g. According to the density functional theory (DFT) calculation,[34] the characteristic peak at ~147 cm$^{-1}$ is related to the out-of-plane mode $A_1^3$ (147 cm$^{-1}$ for $1^{st}$ layer α-Sb), and the one at 131 cm$^{-1}$ to the in-plane mode $A_1^2$ (132 cm$^{-1}$ for $1^{st}$ layer α-Sb), respectively. As the thickness increases, the $A_1^2$ mode at 131 cm$^{-1}$ prominently shifts to 118 cm$^{-1}$ (red shift), while the $A_1^3$ mode at 147 cm$^{-1}$ shifts to 150 cm$^{-1}$ (blue shift). Such Raman shifts can be ascribed to the stacking induced structural change or the anisotropy of the BP-like lattice, similar to what have been observed in black phosphorene and MoS$_2$.[18,46-48] In contrast to the BP-type Bi(110)/HOPG islands which undergo a structural transition at 4 MLs,[36] the multi-layered antimonene always preserves the BP-like structure, and no structural transition is observed. In our STM measurement, the limit of thickness explored is six layers. Raman characterization indicates that the antimonene film keeps the BP puckered structure at least up to ~20 layers (Figure 2g).

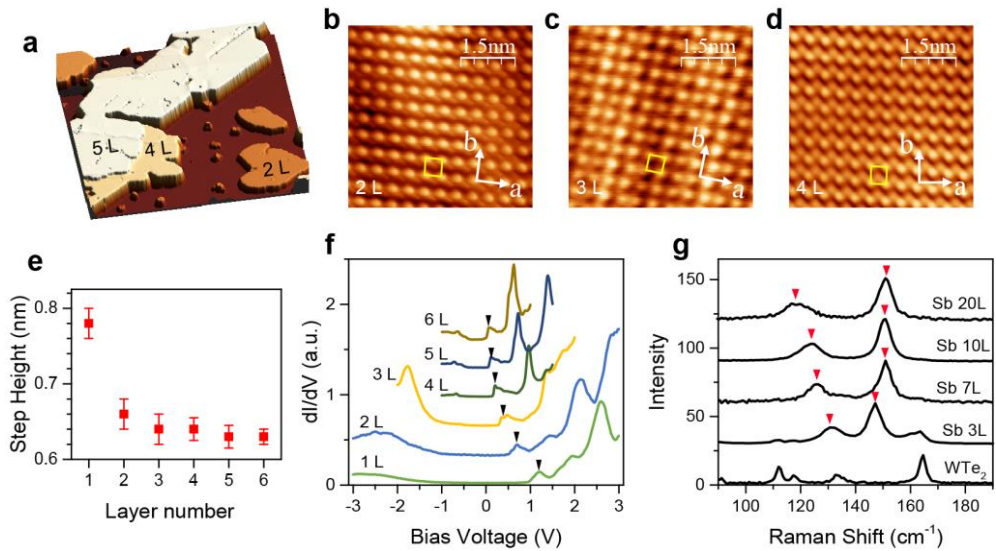

**Figure 2**. Characterization of the epitaxial multi-layered α-antimonene grown on WTe$_2$. a) The STM topographic image (200 × 200 nm$^2$) of multi-layered α-antimonene with thickness varying from 2L to 5L. $U = +2$ V, $I_t = 100$ pA. b-d) Atomically resolved STM images (5 × 5 nm$^2$) obtained on the 2L, 3L and 4L, respectively. $U = +10$ mV, $I_t = 100$ pA. The 1×1 unit cell is marked by the yellow rectangle. e) The step height measured on the α-antimonene films of various thickness. f) Thickness dependent STS taken on the antimonene films from 1L to 6L. The characteristic feature in the unoccupied state is marked by black triangles in each spectrum. The spectra are offset along vertical direction for clarity. g) Raman spectra taken on the bare WTe$_2$ and α-antimonene films with thickness of ~3L, ~7L, ~10L and ~20L, respectively. The spectra are offset along vertical direction for clarity.

The DFT calculated band structure of the antimonene-WTe$_2$ hetero-structure (Figure S6, Supporting Information) reveals that the epitaxial antimonene is hole doped, due to the electron transfer from the antimonene to the underlying WTe$_2$ substrate and as the thickness of Sb increases, the Fermi level gradually shifts upward. Such effects can be confirmed in the following QPI analysis. Regardless of downward shift of the Fermi level, the calculated band structure of the epitaxial antimonene is overall similar with that of the free-standing antimonene, (as shown in Figure S7, Supporting Information), due to the weak interlayer coupling between the antimonene and WTe$_2$ substrate. Therefore, we will use the band structures of freestanding antimonene in the following discussions for simplicity.

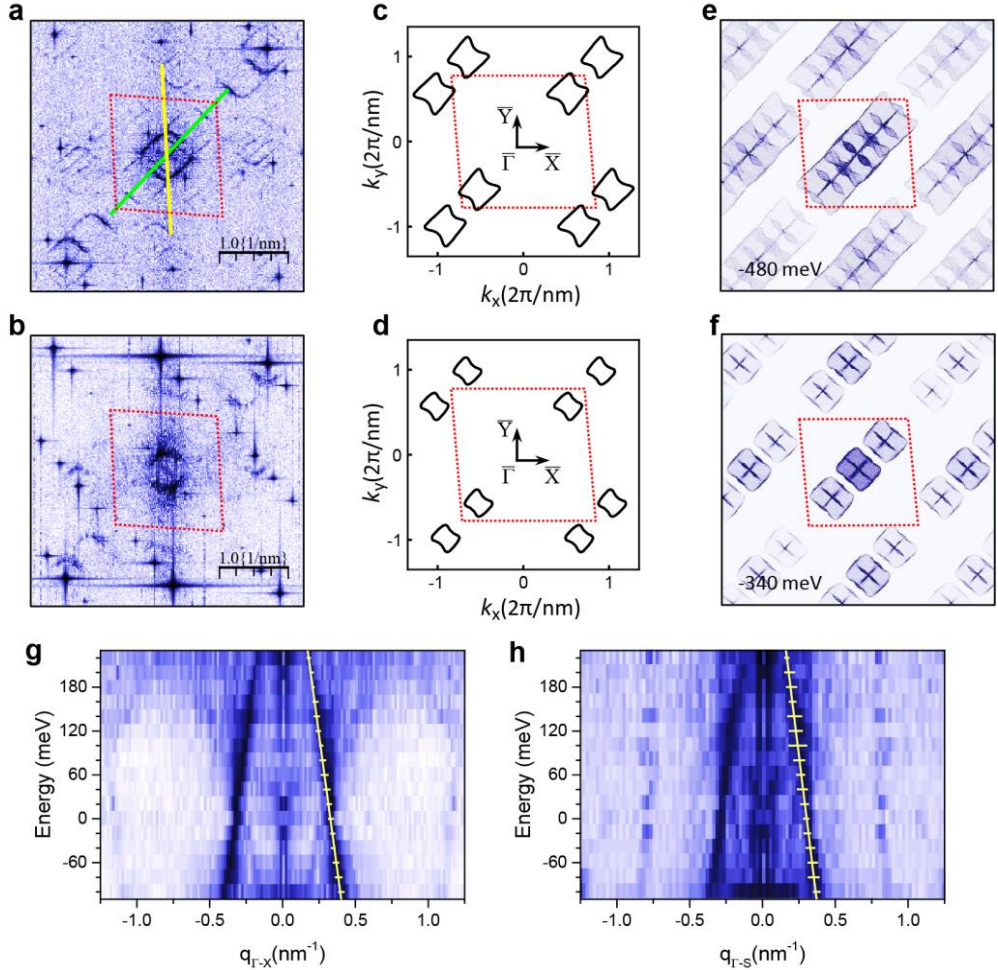

**Figure 3**. Comparison of experimental and DFT simulated QPI patterns and E-q dispersion of single layered α-antimonene. a,b) FFT of experimentally obtained dI/dV maps taken at $U = -20$ mV and $U = +120$ mV. The red dotted square marks the 1$^{st}$ Brillouin Zone (BZ) of R2 reconstruction. c,d) DFT calculated CEC for -480 mV and -340 mV. The high symmetric direction are labeled according to energy band of R2 reconstruction. e,f) The corresponding simulated QPI of (c) and (d). The red square is used to compare with (a) and (b). g,h) E-q dispersion taken along the yellow (Γ-X) and green (Γ-S) lines in (a). The error bars and linear fitting are included in the right side.

To further explore the band structure of antimonene, quasiparticle interference (QPI) is applied to map the dispersive band structure. (**Figure 3**a,b) shows the fast Fourier transform of two QPI patterns measured on the single layer antimonene at U = -20 mV and U = +120 mV respectively (More QPI data can be found in Figure S8 and S9, Supporting Information). Besides the static vectors that are originated from the R2 reconstruction / moiré pattern, the QPI-associated pattern can be identified as the square centered at the Γ point. Along the Γ-S direction (the green line in Figure 3a) are located two replicas of the centered square. The diameter of the square becomes gradually smaller with increasing bias energy, (as shown in Figure S9, Supporting Information), indicating that it is originated from the hole pocket of Sb valence band. The contribution from the WTe$_2$ substrate, as identified as the two arcs symmetrically located along Γ–X [49], is rather weak. Therefore we believe the QPI are dominated by the contribution from the Sb layer. (Figure 3c,d) shows the calculated Constant Energy Contour (CEC) at -480 meV and -340 meV below the valence band maximum for the monolayer α-antimonene in R2 reconstruction. The most dominating pattern is two square-shaped pockets located along the Γ-S direction. Taking into account the hole doping effect, the simulated QPI patterns as shown in (Figure. 3e,f) are qualitatively consistent with the experimental results of (Figure. 3a,b). (Figure 3g,h) shows the E-q dispersion obtained from the line-cuts taken along the Γ-X and Γ-S direction (yellow and green lines in Figure 3a). Both of the E-q curves show the linear dispersion, consistent with the DFT calculated band structure (Figure S7 and S10, Supporting Information). The QPI measurements taken on the bilayer and tri-layer antimonene exhibit more complicated patterns, (as shown in Figure S11 and S12, Supporting Information). Whilst, the linear dispersion crossing the Fermi energy can be still distinguished.

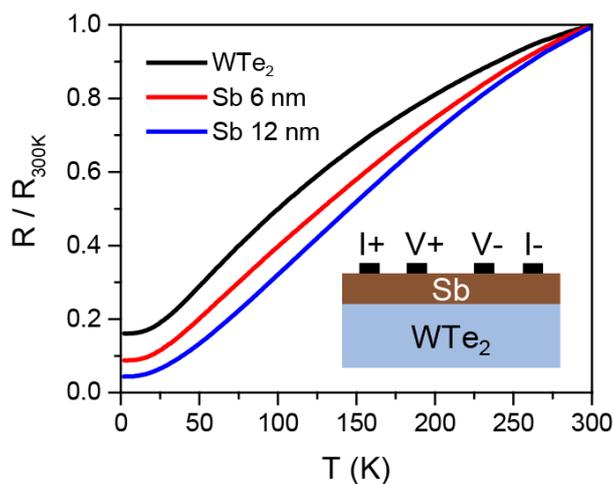

**Figure 4**. Transport measurement of α-antimonene film. Temperature dependence of longitudinal resistance normalized at 300K for bare T$_d$-WTe$_2$ and α-antimonene films with different thickness. The inset shows the schematic of the four-probe resistance measurement.

The agreement of QPI results with the DFT simulations leads to a conclusion that the epitaxial antimonene hold the similar band structure with the free-standing antimonene. The energy band crossing the Fermi level is linearly dispersed, and thus the small effective mass and high carrier mobility are expected, as theoretically predicted.[50] Simply assuming the isotropic case, the electrical conductivity can be described by the formula of $\sigma=ne\mu$, where $\mu$ is inversely proportional to the effective mass. The small effective mass means a high electrical conductivity. To confirm the high electrical conductivity, we measure the T dependent resistance (R) of α-antimonene films, as plotted in **Figure 4**. Referring to the bare $WTe_2$, the resistance of $WTe_2$ coated with ~10 layers of antimonene film is decreased to nearly a half at the low temperature end. As the antimonene film is grown thicker to ~20 Layers, the resistance is further decreased to about a quarter. Even though it is not possible to extract the resistance of pure antimonene due to the conductive $T_d$-$WTe_2$ substrate, the qualitative analysis suggests a good electrical conductivity in α-antimonene film.

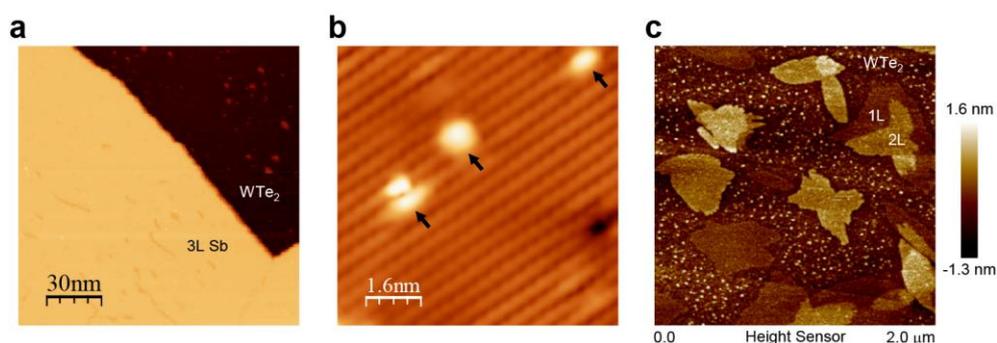

**Figure 5**. Chemical stability of α-antimonene films upon exposure to $O_2$ and air. a) The surface morphology (150 × 150 $nm^2$) of multi-layered antimonene after exposure to pure oxygen gas for ~12000 Langmuir. $U = +2$ V, $I_t = 100$ pA. b) Atomically resolved STM image (8.0 × 8.0 $nm^2$) of the α-antimonene surface. U = +20 mV, $I_t = 200$ pA. The adsorbates are marked by black arrows. c) AFM image taken on the sample after exposure in air for about 12 hours.

Stability is another essential concern considering its practical applications.[46] We investigate the chemical stability of the antimonene film upon exposure to oxygen and air. (**Figure 5**a,b) shows the multi-layered antimonene surface after exposure to oxygen gas of ~12000 Langmuir (~20 min). Other than some adsorbates adsorbed at the surface defective sites, as marked by the arrows in Figure 5b, the surface looks almost the same as before oxygen dosing, Figure 2a. The α-antimonene films are also stable after exposed in air. Figure 5c shows the atomic force microscopy (AFM) image taken after the sample is exposed in air for ~12 hours. The antimonene islands are still kept flat and clean. In contrast, the exposed $WTe_2$ substrate are covered by small clusters, which may be caused by the surface adsorption.

In summary, the high-quality and large-scale α-antimonene with puckered BP structure has been experimentally realized, thus confirming that it is energetically stable. The linearly dispersed metallic band leads to the high electrical conductivity. Such advanced properties, together with its chemical inertness under ambient condition and the structural similarity to BP, make it potential in 2D material-based applications. It is also expected that this study will stimulate further investigation regarding its topological properties or thermal electric applications.

**Experimental Section**

*Sample growth and STM characterization*: The high quality antimonene films were grown in a combined STM-MBE system (Unisoku USM1500) with a base pressure of $1\times10^{-11}$ Torr. The single crystal $T_d$-$WTe_2$ substrate was firstly in situ cleaved in ultrahigh vacuum. The atomic quality of cleaved surface was checked by STM prior to antimonene growth. High purity antimony (alfa Aesar 99.9999%) was evaporated from a standard Knudsen cell. The Sb flux was kept at ~0.3 monolayers per minute. During the growth, the substrate was kept at ~400 K. After growth, the antimonene film was transferred to low-temperature STM stage for characterization. All the STM experiments were performed at 4K with a PtIr tip unless particularly stated. The topographic images were acquired using a constant current mode. The scanning tunneling spectroscopy were acquired with a lock-in amplifier technique. The data were processed with the WSXM software.[51] Oxygen gas was introduced into UHV through a leak valve. The oxygen pressure was kept at $1\times10^{-5}$ Torr during dosing.

*DFT Calculations*: Our first-principles calculations were based on density functional theory (DFT) by using the Vienna Ab Initio Simulation Package (VASP).[52] The projector-augmented wave (PAW) method[53, 54] with a plane wave basis set was used to describe the ion-electron interactions. The exchange-correction potentials were approximated by the gradient approximation (GGA) parameterized by the Perdew-Burke-Ernzerhof (PBE)[55] method. The 2D crystals were modeled by supercells repeated periodically in the 2D plane, while a vacuum region of more than 15 Å was applied to avoid the interactions between the periodic images. Calculations were performed with an energy cutoff of 300 eV on a 12×12×1 Γ-centered Monkhorst-Pack[56] k-point mesh. During structural optimization, the force convergence criterion was set to 0.01 eV/Å. The van der Waals corrections were treated by the semi-empirical DFT-D3 method[57] in the bilayer and trilayer calculations.

To investigate the electronic structure information from the QPI patterns, we simulated the FT-STS images using the joint density of states (JDOS) approximation, based on the self-correlation function of the 2D constant energy contours (CEC) of the surface states at a given energy.[58, 59] The factors of the scattering events between different surface states were simply set to be a constant.

*Raman, AFM and Resistance Measurement*: Raman spectra were performed using a Witec/alpha 300 R confocal microscope with a 532 nm laser at ambient conditions, and AFM images were taken using the Bruker AXS Dimension Icon in tapping mode. The electrical resistances of Sb films and $WTe_2$ substrate were measured in a physical property measurement system (PPMS, Quantum Design).

**Supporting Information**

Supporting Information is available from the Wiley Online Library or from the author.

**Acknowledgements**

This work was financially supported by the Ministry of Science and Technology of China (Grants Nos. 2014CB921103, 2017YFA0204904, 2015CB921203), the National Natural Science Foundation of China (Grants Nos. 11774149, 11790311, 11674299, 11634011, 11674154, 11374149, 11374140), the Strategic Priority Research Program of Chinese Academy of Sciences, Grant No. XDB30000000, the Fundamental Research Funds for the Central Universities (Grant Nos. WK2340000063, WK2340000082, WK2060190084), and the Open Research Fund Program of National Laboratory of Solid State Microstructures. Computational support was provided by National Supercomputing Center in Tianjin.